\documentclass[a4paper,11pt]{article}
\usepackage{pos}

\makeatletter
\g@addto@macro\bfseries{\boldmath}
\makeatother

\title{Nucleon-nucleon scattering from distillation}

\renewcommand{\logo}{\relax} 

\manuallySeparateAuthors 

\author*[a]{Jeremy~R.~Green,}
\renewcommand{\printHeadAuthors}{Jeremy R.\ Green} 
\author[b]{ Andrew~D.~Hanlon,}
\author[c]{ Parikshit~M.~Junnarkar}
\author[d,e]{ and Hartmut~Wittig}
\author{ for the Baryon Scattering (BaSc) collaboration}

\affiliation[a]{Deutsches Elektronen-Synchrotron DESY,\\
Platanenallee 6, 15738 Zeuthen, Germany}

\affiliation[b]{Physics Department, Brookhaven National Laboratory,\\
Upton, New York 11973, USA}

\affiliation[c]{Institut für Kernphysik, Technische Universität Darmstadt,\\
Schlossgartenstraße 2, 64289 Darmstadt, Germany}

\affiliation[d]{PRISMA Cluster of Excellence and Institut für Kernphysik, University of Mainz,\\
Becher Weg 45, 55099 Mainz, Germany}

\affiliation[e]{GSI Helmholtzzentrum für Schwerionenforschung,\\
64291 Darmstadt, Germany}

\emailAdd{jeremy.green@desy.de}
\emailAdd{ahanlon@bnl.gov}
\emailAdd{parikshit@theorie.ikp.physik.tu-darmstadt.de}
\emailAdd{hartmut.wittig@uni-mainz.de}

\abstract{
  
  We report an ongoing analysis of nucleon-nucleon scattering based on
  finite-volume spectroscopy. The calculation is performed using the
  distillation method on eight lattice ensembles at the
  SU(3)-symmetric point with $m_\pi=m_K\approx 420$~MeV generated by
  CLS, covering a range of lattice spacings and volumes and previously
  used to study the $H$~dibaryon. We obtain nonzero signals for $S$,
  $P$, $D$, and $F$ waves as well as the mixing between spin-1 $S$ and
  $D$ waves. For $S$ waves, lattice artifacts are significant and tend
  to strengthen baryon-baryon interactions. In the deuteron and
  dineutron $S$ waves, we find virtual bound states.

}

\FullConference{%
  The 39th International Symposium on Lattice Field Theory (Lattice2022),\\
  8-13 August, 2022 \\
  Bonn, Germany 
}

\renewcommand{\hookAfterAbstract}{%
  \par\bigskip
  \hfill DESY-22-202
}

\begin{document}
\maketitle

\section{Introduction}

In Ref.~\cite{Green:2021qol}, we reported a study of the $H$~dibaryon
at the SU(3)-symmetric point where the $u$, $d$, and $s$ quark masses
are set to their physical average value. Working at a fixed quark mass
point, we were able to study a wide range of lattice spacings and
multiple volumes. Surprisingly, we found that the binding energy of
the $H$ dibaryon is significantly affected by discretization effects,
ranging from about 5~MeV in the continuum to above 30~MeV on the
coarsest lattice spacing.

In these proceedings, we report work in progress to study
nucleon-nucleon scattering using the same dataset --- at this stage,
all results should be considered preliminary. Concerning the presence
of $NN$ bound states at heavier-than-physical pion masses, there is a
disagreement in the literature~\cite{Iritani:2017rlk}: studies based
on asymmetric point-source correlation functions (with a hexaquark
interpolator at the source and a baryon-baryon interpolator at the
sink) find a bound deuteron and dineutron~\cite{Beane:2012vq,
  Orginos:2015aya, Wagman:2017tmp, NPLQCD:2020lxg, Yamazaki:2012hi,
  Yamazaki:2015asa, Berkowitz:2015eaa}, whereas studies using the
variational method with a symmetric correlator matrix or using
HAL~QCD's potential method find no bound state~\cite{Inoue:2011ai,
  Horz:2020zvv, Nicholson_proc}. We also note the recent variational
calculation from NPLQCD~\cite{Amarasinghe:2021lqa, Wagman_proc} using
ensembles previously employed in Refs.~\cite{Beane:2012vq,
  Wagman:2017tmp, Berkowitz:2015eaa} that appears to be consistent
with \cite{Horz:2020zvv} but did not make a final decision about the
presence of a bound state.

In the next section, we briefly summarize our methodology; in
Section~\ref{sec:spectra}, we show the finite-volume and
nonzero-lattice-spacing spectra; in Section~\ref{sec:Swave}, we study
the $S$-wave phase shifts while neglecting higher partial waves; in
Section~\ref{sec:higher}, we show some higher partial waves; and in
Section~\ref{sec:mixing}, we analyze the mixing between $S$ and $D$
waves. Finally, we give our conclusions.

\section{Lattice setup}\label{sec:lattice}

The details of our calculation are the same as in
Ref.~\cite{Green:2021qol}. Matrices of two-point correlation functions
were computed using the distillation method~\cite{Peardon:2009gh} on
eight ensembles with nonperturbatively $O(a)$-improved Wilson-clover
fermions generated by CLS~\cite{Bruno:2014jqa} with
$m_\pi=m_K\approx 420$~MeV, spanning six lattice spacings from 0.039
to 0.099~fm and $m_\pi L$ varying between 4.4 and 6.4.

Our analysis of the spectra is based on finite-volume quantization
conditions~\cite{Luscher:1990ux, Rummukainen:1995vs, Briceno:2013lba,
  Briceno:2014oea}. Roughly following Ref.~\cite{Morningstar:2017spu},
the finite-volume spectrum is given by solutions of
\begin{equation}
  \det\left[ \tilde K^{-1}(p^2) - B(p^2) \right] = 0,
\end{equation}
where $\tilde K$ contains the scattering amplitude and $B$ depends on
the volume, $\vec P$, and irreducible representation of the little
group of $\vec P$. In this work, our preferred kinematic variable is
the centre-of-mass momentum $p^2\equiv (E_\text{cm}/2)^2-m^2$. Given
an ansatz for $\tilde K^{-1}(p^2)$, we find the solutions $\{p^2\}$
and compare them with the observed spectrum, performing a correlated
least-squares minimization.

Here we study the flavour septenvigintuplet, which lies in the
symmetric product of two octets and contains $NN$ $I=1$, and the
antidecuplet, which lies in the antisymmetric product and contains
$NN$ $I=0$. Our interpolating operators are formed from linear
combinations of products of momentum-projected single-baryon
interpolators and have definite flavour, total momentum $\vec P$,
irrep, and two-baryon spin. Although spin is not a preserved quantum
number, both the scattering amplitude for identical baryons and the
two-particle finite-volume quantization condition are diagonal in
spin, implying that spin zero and spin one can be analyzed
separately. Furthermore, we observe that every state overlaps largely
with operators with only one spin, allowing us to sort the states into
spin zero or one.

\section{Nucleon-nucleon spectra}\label{sec:spectra}

\begin{figure}
  \centering
  \includegraphics[width=\textwidth]{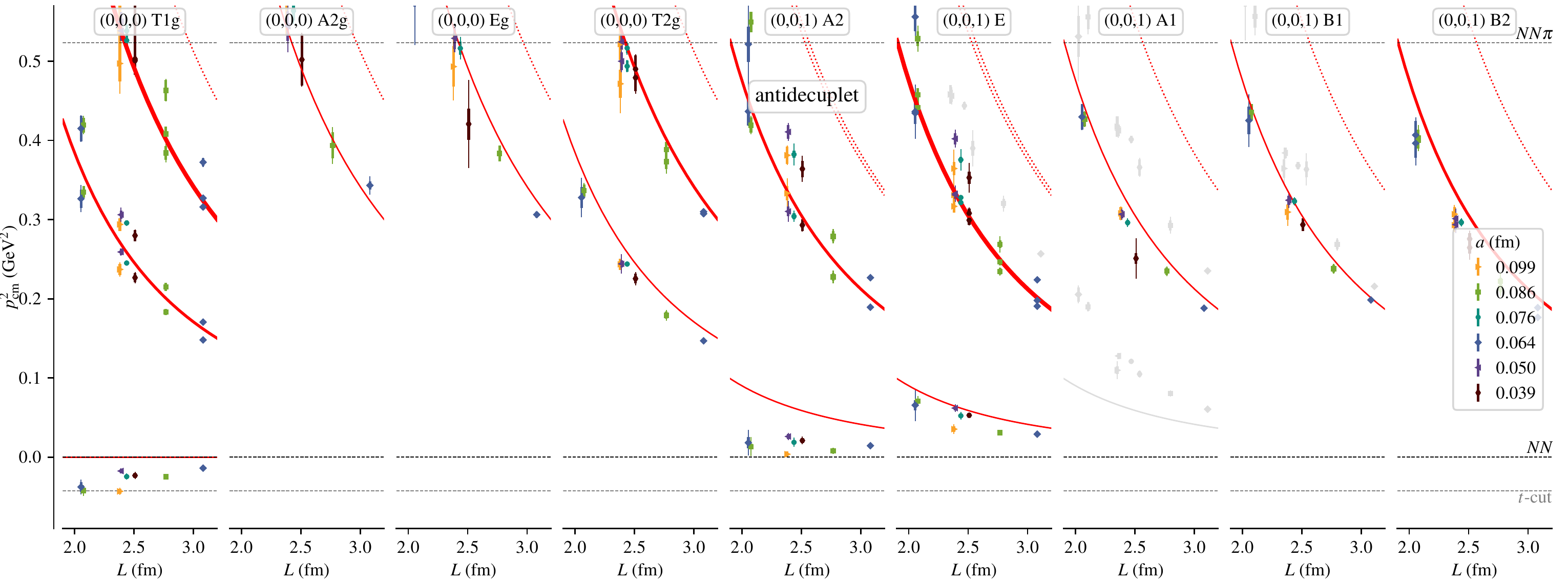}
  \includegraphics[width=\textwidth]{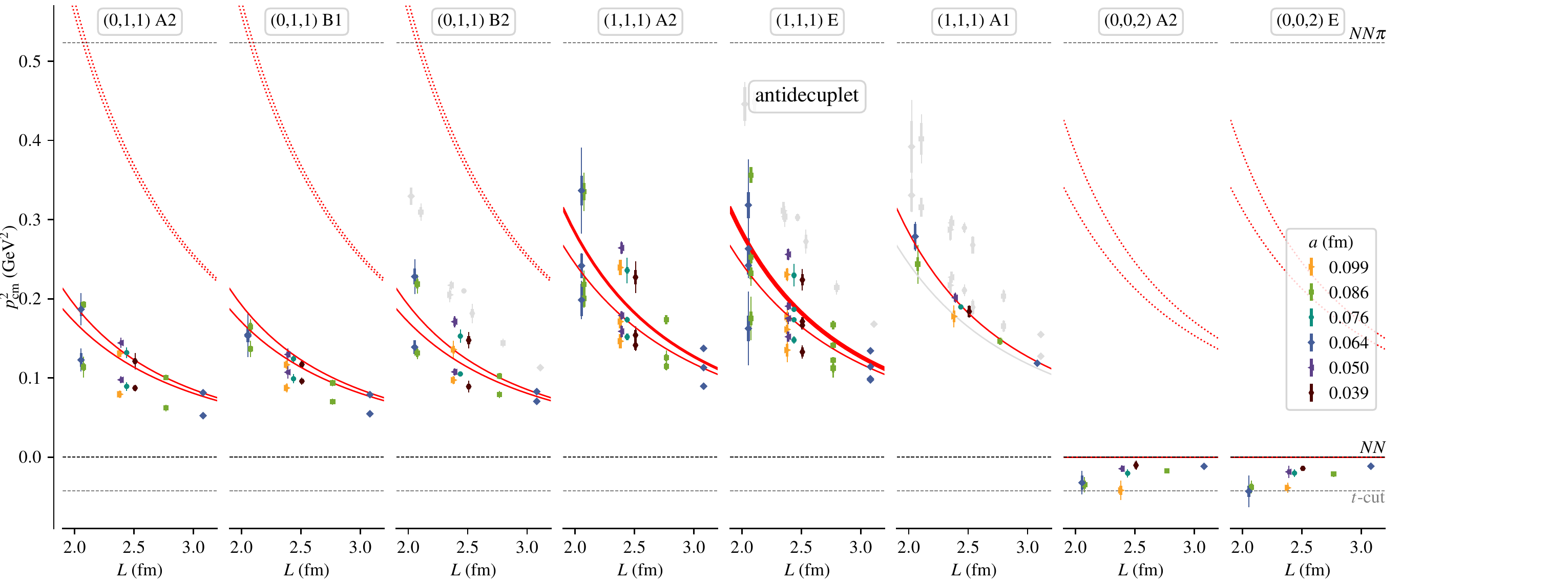}
  \caption{Antidecuplet ($NN$ $I=0$) spin-one spectrum for various
    total momenta and irreps: $p^2$ versus $L$. Red curves show
    noninteracting levels with a thickness proportional to the
    degeneracy; dashed curves show the lowest levels for which an
    interpolating operator was not included. Points correspond to
    lattice energy levels, with the thin outer error bar including an
    estimate of systematic uncertainty based on varying the plateau
    fit range. Gray points indicate levels that are identified as spin
    zero. Horizontal lines give the locations of thresholds and the
    $t$-channel cut.}
  \label{fig:spectra}
\end{figure}

We perform fits to ratios between diagonalized two-baryon correlators
and the product of two single-baryon correlators to obtain energy
differences from noninteracting levels. The systematic uncertainty is
estimated using an alternative fit range. When we fit to the spectra to
determine scattering amplitudes, currently we follow the same approach
as Ref.~\cite{Green:2021qol} and treat this systematic as fully
correlated, which means that it tends not to significantly reduce
$\chi^2$; in the future, we may decide to change this. Also following
Ref.~\cite{Green:2021qol}, we use bootstrap to estimate statistical
errors of the scattering parameters and fit the alternative spectra to
estimate part of the systematic uncertainty.

A large number of energy levels is obtained. For example, the
isospin-zero spin-one case is shown in Fig.~\ref{fig:spectra}, where
we obtain over 300 levels. Fitting these will be a challenge and will
require at minimum the ${}^3S_1$, ${}^3D_1$, ${}^3D_2$, and ${}^3D_3$
phase shifts\footnote{We denote partial waves as ${}^{2s+1}\ell_J$,
  where $s$ is the two-baryon spin and $\ell$ indicates the orbital
  angular momentum.} together with the mixing angle for $J=1$, all of
which depend on $p^2$ and may also be affected by lattice artifacts.

For many energy levels, a pattern is visible across the different
ensembles similar to what was observed for the $H$~dibaryon: coarser
ensembles produce lower-lying energies, generally corresponding to a
stronger attraction. The long-sought-after splitting between the
ground states in the A2 and E irreps in frame $(0,0,1)$ is also
evident: this is a clear signal of mixing between ${}^3S_1$ and
${}^3D_1$~\cite{Briceno:2013bda}.

\section{$S$-wave phase shifts}\label{sec:Swave}

We begin by neglecting $D$-wave contributions and choosing energy
levels to minimize the influence of higher partial waves. Under this
approximation, each energy level directly yields the $S$-wave phase
shift at that energy.

\begin{figure}
  \centering
  \includegraphics[width=0.7\textwidth]{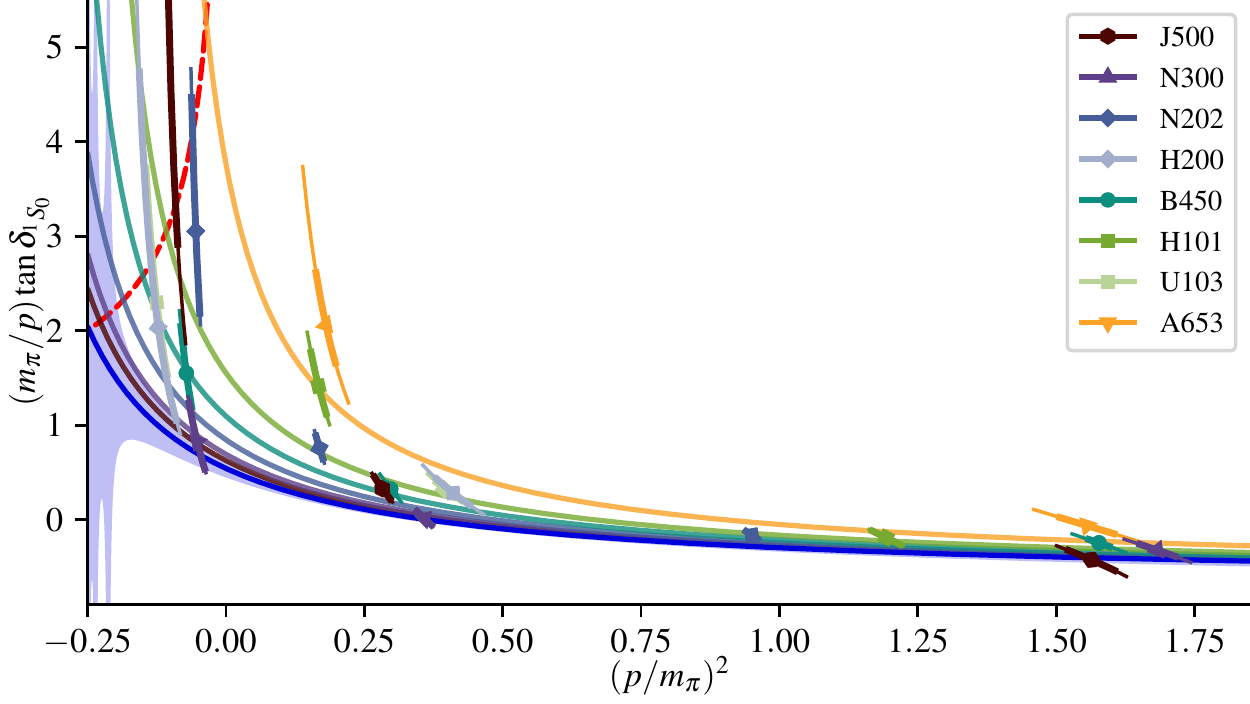}
  \caption{Septenvigintuplet ($NN$ $I=1$) ${}^1S_0$ phase shift:
    $p^{-1}\tan\delta$ versus $p^2$, normalized with the pion
    mass. Darker and more purple points and curves correspond to finer
    lattice spacings and pale points indicate the two small-volume
    ensembles. The blue curve with error band is the continuum limit
    of the fit. Intersections with the red dashed curve correspond to
    virtual state poles.}
  \label{fig:I1Swave}
\end{figure}

For the 27-plet ($NN$ $I=1$), we observe that the phase shift passes
through zero above threshold while becoming large below
threshold. This leads us to use a rational function as the fit ansatz,
$p\cot\delta(p)=(c_0+c_1 p^2)/(1+c_2 p^2)$, where $c_i$ are affine
functions of $a^2$. We select the ground and first-excited state in
the rest frame irrep A1g, along with the ground state in the first
moving frame irrep A1. On the two small-volume ensembles, we exclude
the rest-frame excited state, since it lies above the $N\Delta$
threshold. The fit produces a rough description of the data, but
quantitatively it is not particularly good, with
$\chi^2/\text{dof}=23/16$. The data and the fit are shown in
Fig.~\ref{fig:I1Swave}. The phase shift tends to decrease as the
continuum limit is approached. There is a virtual state pole, which
moves further below threshold in the continuum.

\begin{figure}
  \centering
  \includegraphics[width=0.7\textwidth]{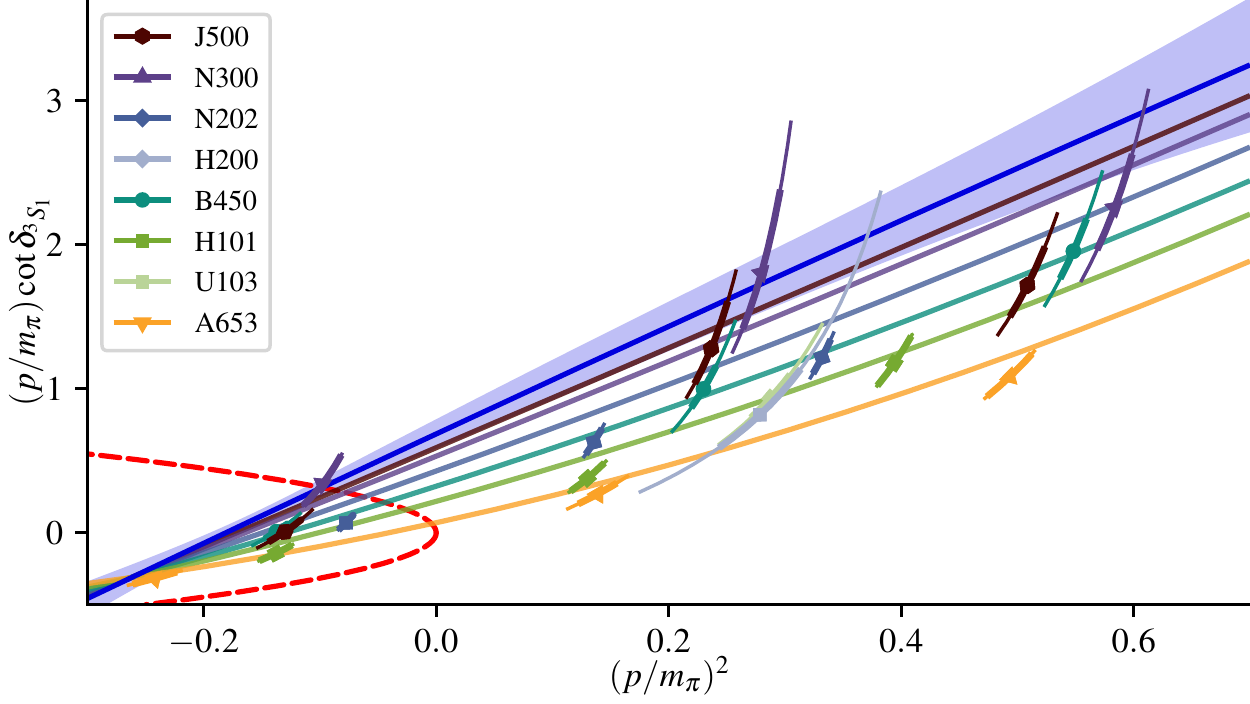}
  \caption{Antidecuplet ($NN$ $I=0$) ${}^3S_1$ phase shift, in the
    approximation that the ${}^3D_1$ partial wave and mixing with it
    vanish. See the caption of Fig.~\ref{fig:I1Swave}. An intersection
    the red dashed curve corresponds to a pole in the scattering
    amplitude: a virtual state if on the upper branch and a bound
    state if on the lower branch.}
  \label{fig:I0Swave}
\end{figure}

For the antidecuplet ($NN$ $I=0$), the presence of mixing between
${}^3S_1$ and ${}^3D_1$ partial waves complicates the analysis. For a
first attempt, in the first and second moving frames we take the
helicity-averaged ground-state levels~\cite{Briceno:2013bda}. In frame
$(0,0,1)$, this corresponds to averaging the energy levels in irreps E
and A2 with weights 2 and 1 to account for the fact that two of three
helicities lie in the E irrep. We then treat the resulting averaged
levels as arising from a purely $S$-wave interaction. We use a
quadratic polynomial in $p^2$ with coefficients that are affine
functions of $a^2$ as our fit ansatz for $p\cot\delta(p)$, which again
yields worse-than-desired fit quality but still roughly describes the
data: $\chi^2/\text{dof}=32/14$. This is shown in
Fig.~\ref{fig:I0Swave}; as in previous cases, we find that going to
finer lattice spacings produces a weaker interaction. For our coarsest
lattice spacing, there is perhaps a bound state at threshold, but this
turns into a virtual state and moves below threshold in the continuum.
Note, however, that the usefulness of the helicity-averaged
approximation is demonstrated empirically in
Ref.~\cite{Briceno:2013bda} based on experimentally measured
scattering amplitudes, and its validity should likewise be checked for
the setup used here. We also note that our lowest-lying levels are
very close to the $t$-channel cut, where existing quantization
conditions are not valid; an approach that accounts for the leading
$t$-channel exchange was presented at this
conference~\cite{Raposo_proc}.

\section{Higher partial waves}\label{sec:higher}

Our data are also sensitive to higher partial waves. The simpler cases
are those that do not mix, i.e.\ those with orbital and total angular
momentum equal ($\ell=J$) as well as ${}^3P_0$. For the corresponding
$P$ and $D$ waves, there always exists at least one frame and irrep
for which no other equal or lower partial wave contributes. Thus,
neglecting higher partial waves, each energy level in these irreps
yields the corresponding phase shift at that energy
level.

\begin{figure}
  \centering
  \includegraphics[width=0.32\textwidth]{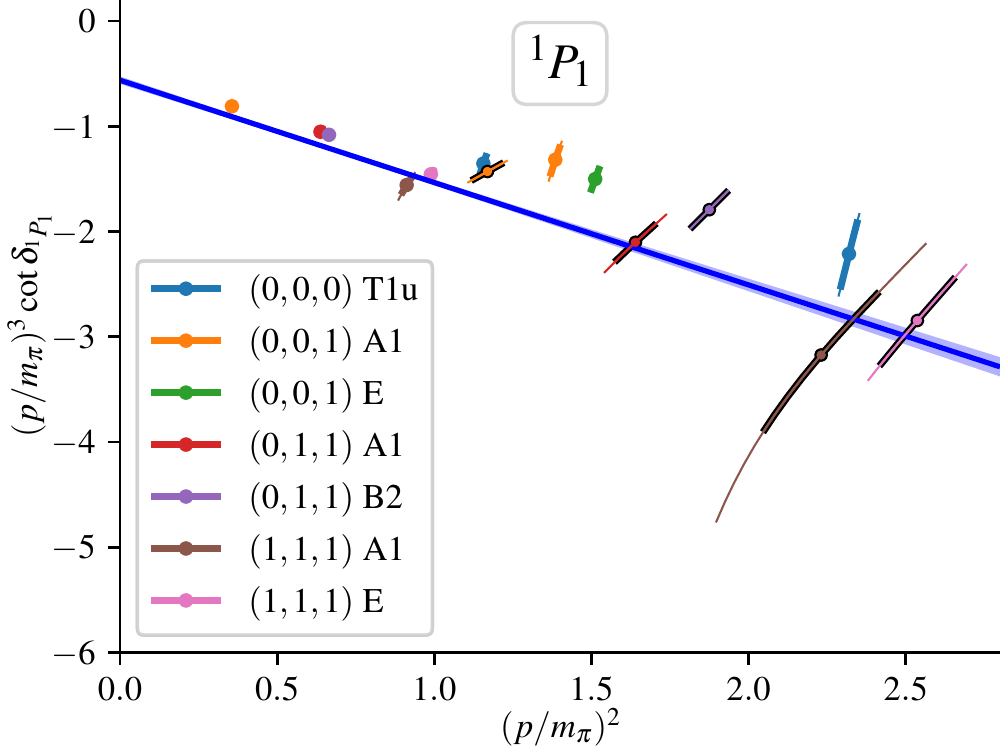}
  \includegraphics[width=0.32\textwidth]{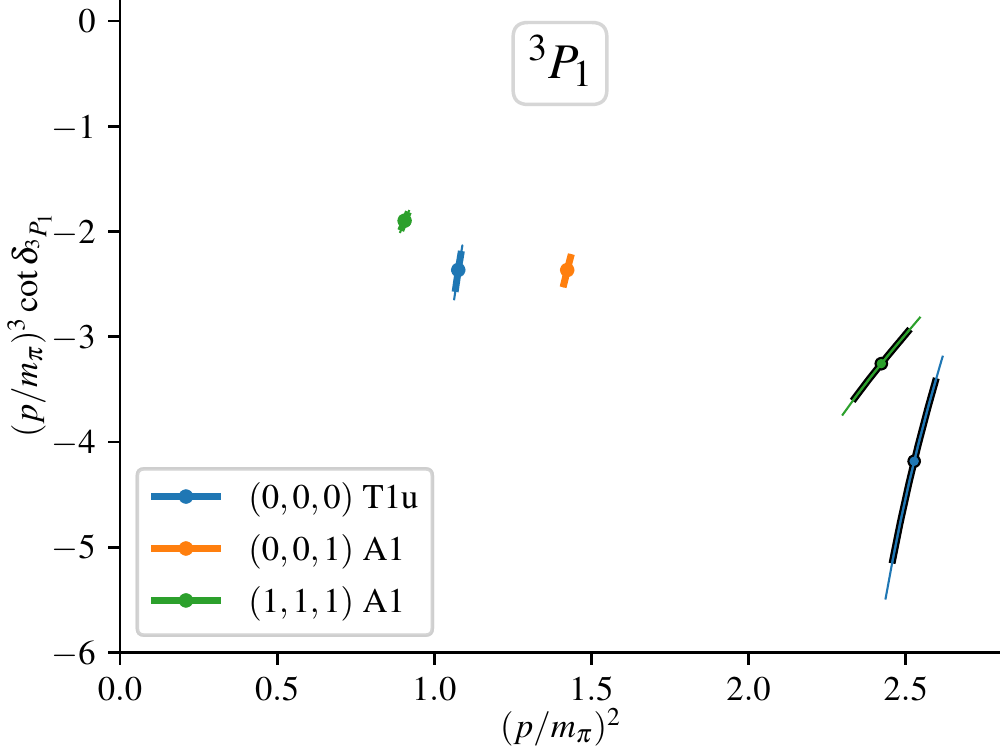}
  \includegraphics[width=0.32\textwidth]{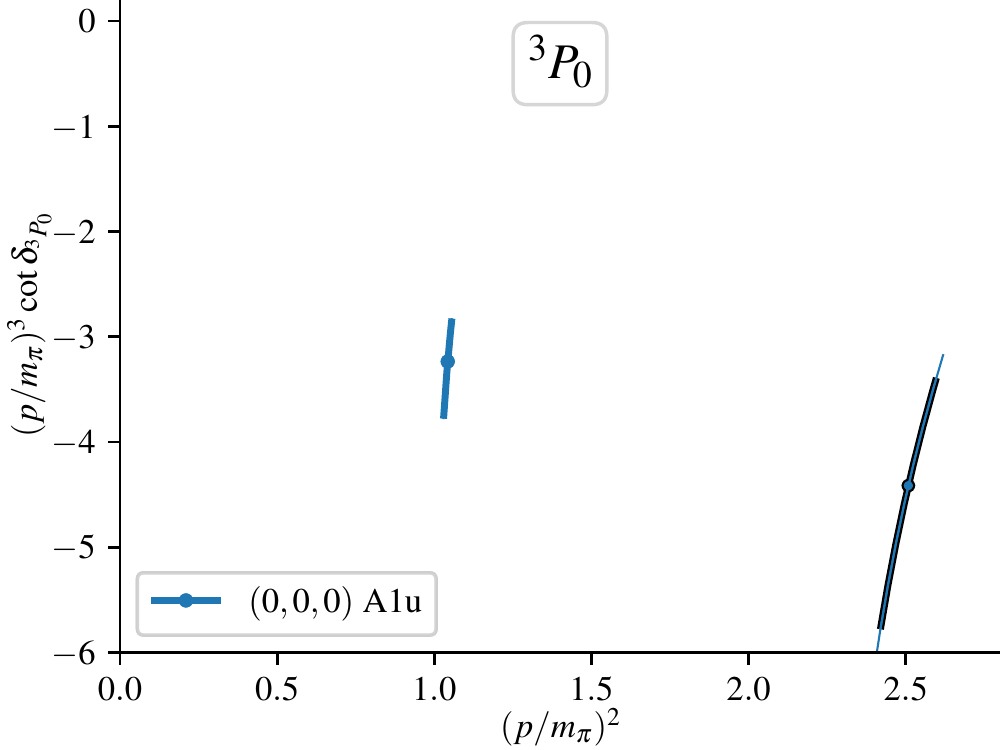}\\
  \includegraphics[width=0.32\textwidth]{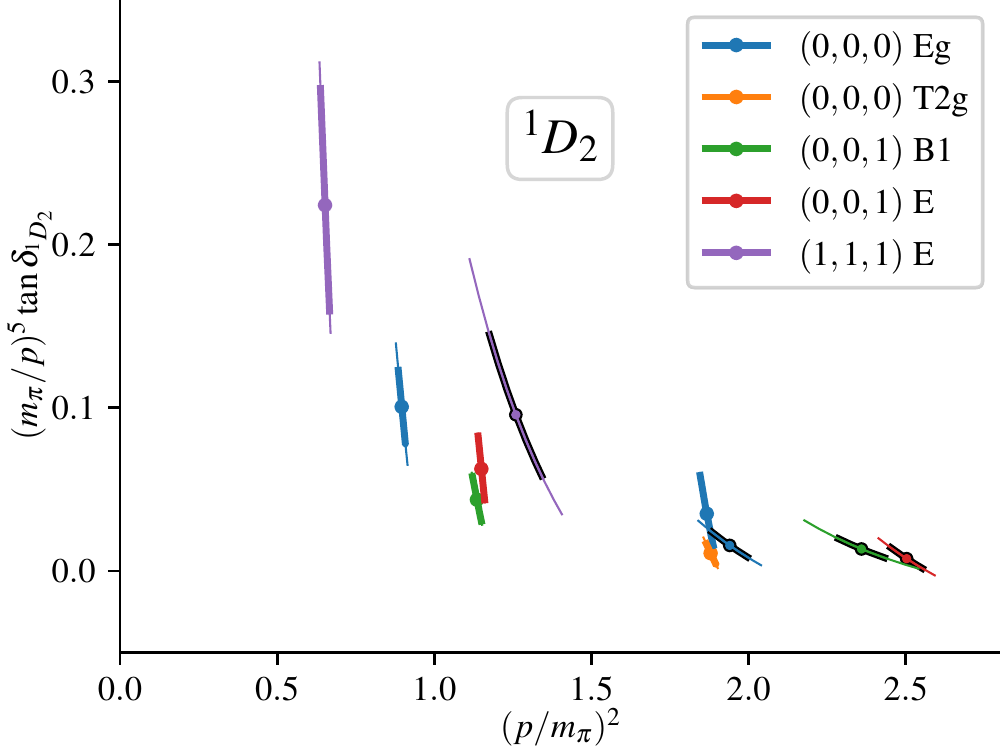}
  \includegraphics[width=0.32\textwidth]{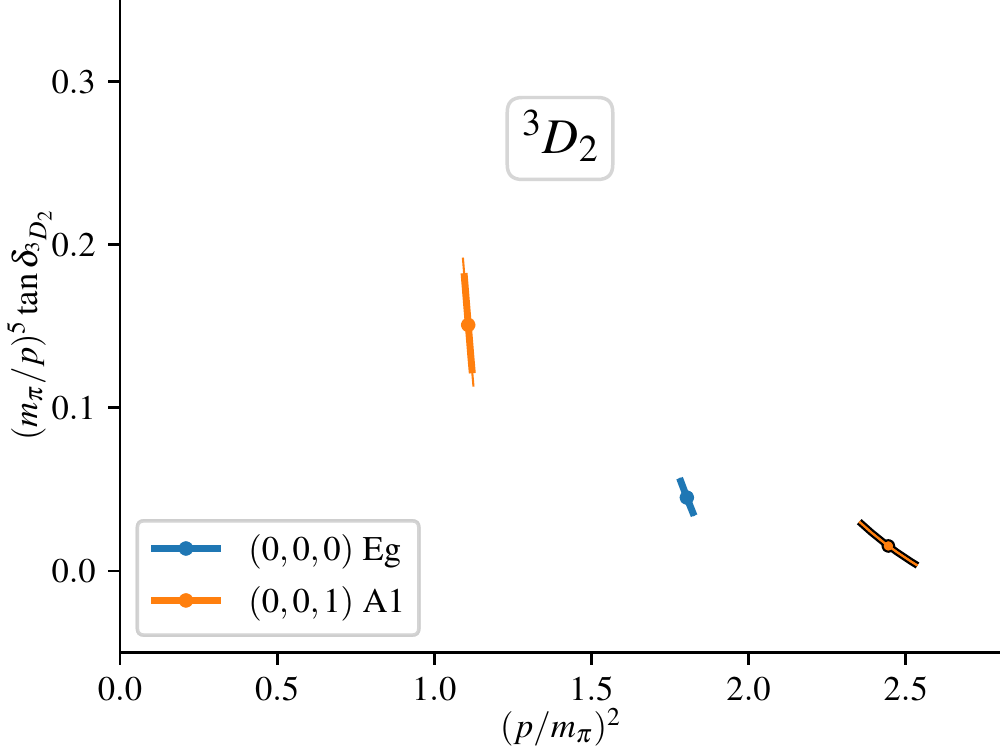}
  \includegraphics[width=0.32\textwidth]{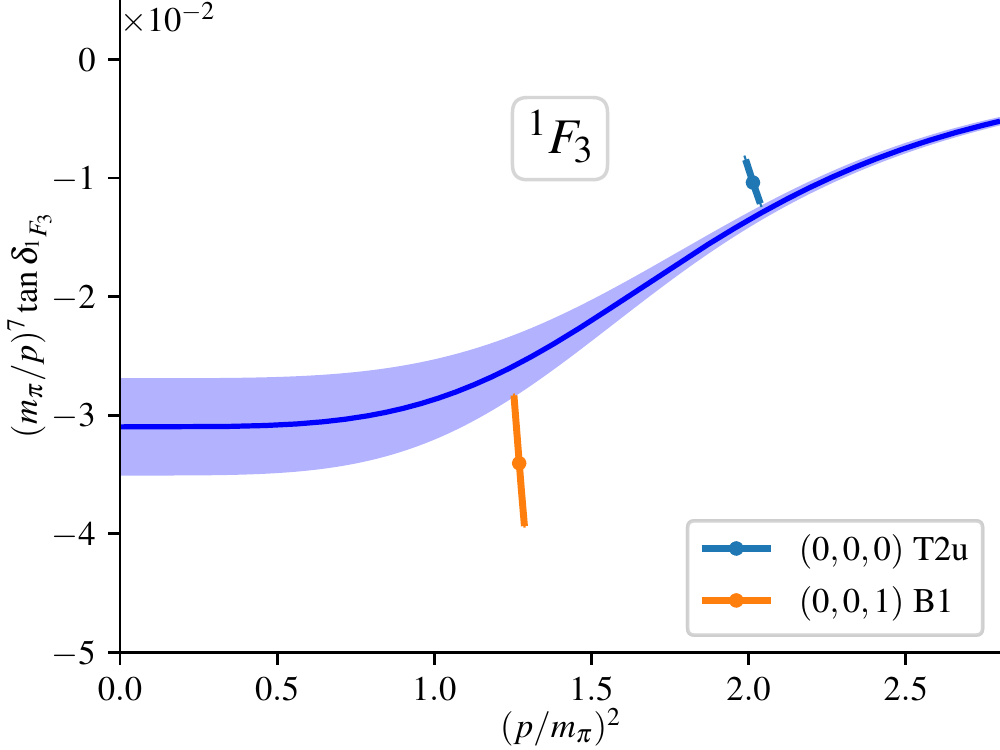}
  \caption{Higher partial waves from ensembles N202 ($L/a=48$, solid)
    and H200 ($L/a=32$, with black outline) with $a=0.064$~fm. First
    row: $P$ waves, $p^3\cot\delta$ versus $p^2$. Second row: $D$ and
    $F$ waves, $p^{-(2\ell+1)}\tan\delta$ versus $p^2$. The ${}^1P_1$,
    ${}^3D_2$, and ${}^1F_3$ partial waves have $I=0$, while
    ${}^1D_2$, ${}^3P_1$, and ${}^3P_0$ have $I=1$. For ${}^1P_1$, we
    have omitted the lowest-lying level in $(1,1,1)$ A1, which is more
    strongly influenced by the ${}^1F_3$ phase shift. The blue curves
    show the fit to all ensembles depicted in Fig.~\ref{fig:PFfit}.}
  \label{fig:higher_partial_waves}
\end{figure}

Figure~\ref{fig:higher_partial_waves} shows the data for two ensembles
at a single lattice spacing. For the spin-zero cases, many energy
levels are obtained, which yields a clear picture of the phase
shift. In the spin one cases, the increased number of partial waves
reduces the number of levels useful in this way, but we still obtain a
good number of constraints on the phase shifts.

\begin{figure}
  \centering
  \includegraphics[width=\textwidth]{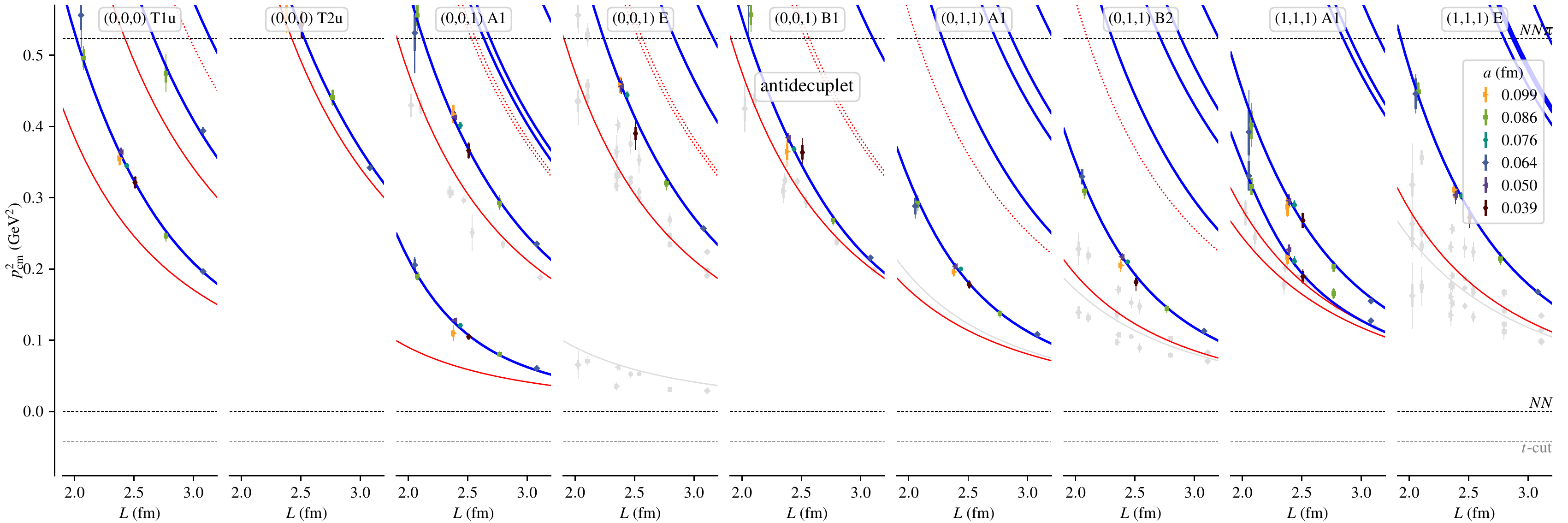}
  \caption{Antidecuplet ($NN$ $I=0$) spin-zero spectrum. See the caption of Fig.~\ref{fig:spectra}. The blue curves show the fit described in the text.}
  \label{fig:PFfit}
\end{figure}

For a full analysis of the spectrum, levels that provide information
about more than one partial wave can also be included, which will
yield additional constraints on the phase shifts. An example fit for
${}^1P_1$ and ${}^1F_3$ is shown in
Figs.~\ref{fig:higher_partial_waves} and \ref{fig:PFfit}: we use the
four-parameter ansatz
\begin{equation}
  p^3\cot\delta_{{}^1P_1} = c_0 + c_1 p^2,\qquad
  p^7\cot\delta_{{}^1F_3} = c_2 + c_3 p^8,
\end{equation}
which is designed to make $\delta_{{}^1F_3}$ go to zero at large
$p^2$. Assuming no discretization effects, we obtain a good fit to all
ensembles with $\chi^2/\text{dof}=51/72$. This suggests that lattice
artifacts may be less relevant for higher partial waves. Here it was
essential to include the $F$ wave; neglecting this is partly
responsible for the disagreement between the curve and the points for
${}^1P_1$ in Fig.~\ref{fig:higher_partial_waves}.

\section{${}^3S_1$--${}^3D_1$ partial wave mixing}\label{sec:mixing}

For the coupled ${}^3S_1$ and ${}^3D_1$ partial waves, we use the
Blatt-Biedenharn parametrization~\cite{Blatt:1952zza},
\begin{equation}
    \tilde K^{-1} =
    \begin{pmatrix}
      1 & 0 \\ 0 & p^2
    \end{pmatrix}
    \begin{pmatrix}
      \cos\epsilon_1 & -\sin\epsilon_1 \\
      \sin\epsilon_1 &  \cos\epsilon_1
    \end{pmatrix}
    \begin{pmatrix}
      p\cot\delta_{1\alpha} & 0\\
      0 & p\cot\delta_{1\beta}
    \end{pmatrix}
    \begin{pmatrix}
      \cos\epsilon_1 &  \sin\epsilon_1 \\
     -\sin\epsilon_1 &  \cos\epsilon_1
   \end{pmatrix}
    \begin{pmatrix}
      1 & 0 \\ 0 & p^2
    \end{pmatrix},
\end{equation}
which encodes the diagonalization of the S-matrix. In the limit of
zero mixing angle $\epsilon_1$, the $\alpha$ wave corresponds to
${}^3S_1$ and the $\beta$ wave to ${}^3D_1$. Using the approximation
$\delta_{1\beta}=0$, each energy level imposes a constraint on the
$(p^{-2}\tan\epsilon_1,p\cot\delta_{1\alpha})$ plane:
\begin{equation}
      p\cot\delta_{1\alpha} = \frac{B_{00} + (B_{01}+B_{10})x + B_{11}x^2}{1+p^4x^2},\quad
      x = p^{-2}\tan\epsilon_1,  
\end{equation}
where $B_{ij}$ is the finite-volume matrix $B(p^2)$.

\begin{figure}
  \centering
  \includegraphics[width=0.68\textwidth]{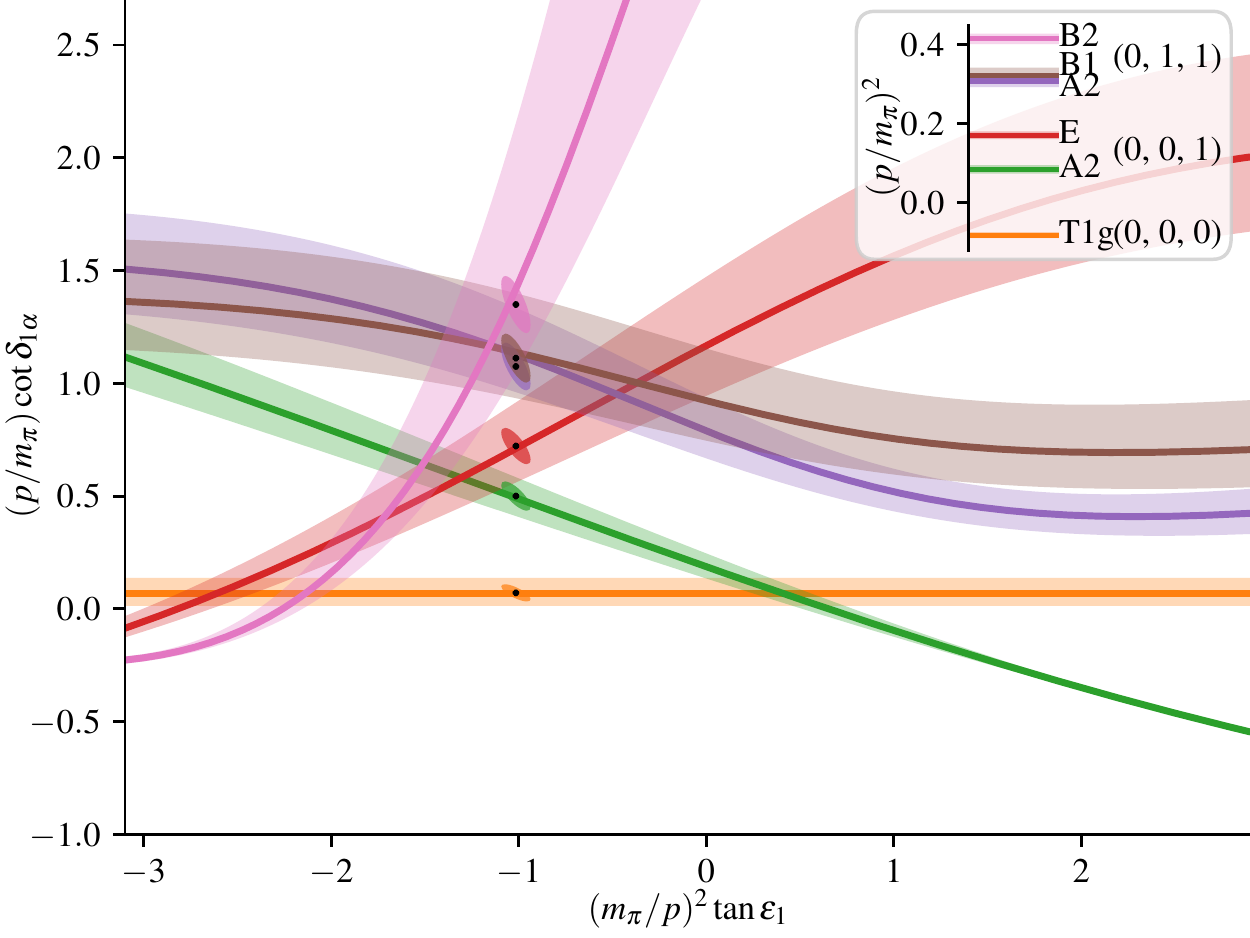}
  \includegraphics[width=0.3\textwidth]{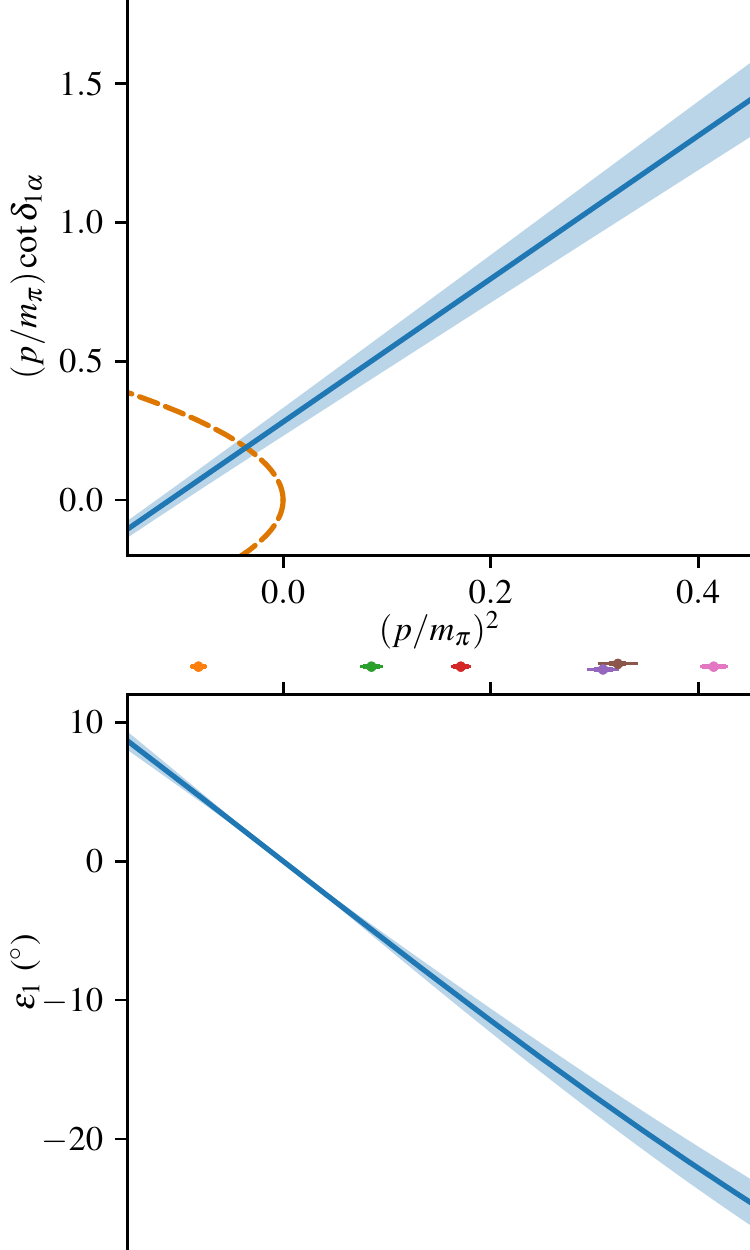}
  \caption{Ensemble N202: $\alpha$-wave phase shift and $J=1$ mixing
    angle, assuming $D$-wave phase shifts vanish. Left: constraints
    imposed by the different energy levels, with inset indicating the
    $p^2$ values. The points with error ellipses show the fit,
    evaluated at the corresponding $p^2$. Right: fitted
    $p\cot\delta_{1\alpha}$ and $\epsilon_1$ versus $p^2$.}
  \label{fig:mixing}
\end{figure}

For ensemble N202 (our largest volume), these constraints are shown
for ground states in the rest frame and first two moving frames in
Fig.~\ref{fig:mixing}. In the moving frames, different irreps provide
quite different constraints and there is sensitivity to the mixing
angle. We fit these six energy levels with the simple three-parameter
ansatz
\begin{equation}
  p\cot\delta_{1\alpha} = c_1 + c_2 p^2,\qquad
  p^{-2}\tan\epsilon_1 = c_3,
\end{equation}
obtaining $\chi^2/\text{dof}=1.2/3$, and the fit results are also shown
in the figure. Like for the analysis using the helicity-averaged
approximation, we find a virtual state pole. Above threshold, the
mixing angle is negative, which should be contrasted with the positive
mixing angle observed in experiments~\cite{Briceno:2013bda}.

\section{Conclusions}\label{sec:conclusions}

Using the distillation method, we were able to employ many
interpolating operators and obtain a large number of finite-volume
nucleon-nucleon energy levels. Similar to what was found in our study
of the $H$ dibaryon, discretization effects appear to be significant
also for the $NN$ spectrum. With the action used by CLS, lattice
artifacts tend to strengthen $S$-wave baryon-baryon interactions. It
appears that large lattice artifacts will add to the numerous existing
challenges for computing the physical deuteron's small binding energy.

We find virtual state poles in the $I=0$ and $I=1$ $NN$ $S$ waves at
the SU(3)-symmetric point. Together with the opposite sign for the
mixing angle $\epsilon_1$, this suggests that lighter pion masses will
be needed to connect with the physics of the deuteron.

\small
\acknowledgments

We thank Raúl A.\ Briceño
for a helpful conversation.
Calculations for this project used resources on the supercomputers
JUQUEEN~\cite{juqueen}, JURECA~\cite{jureca}, and JUWELS~\cite{juwels}
at Jülich Supercomputing Centre (JSC). The authors gratefully
acknowledge the support of the John von Neumann Institute for
Computing and Gauss Centre for Supercomputing e.V.\
(\url{http://www.gauss-centre.eu}) for project HMZ21.
The raw distillation data were computed using
QDP++~\cite{Edwards:2004sx}, PRIMME~\cite{PRIMME}, and the deflated
SAP+GCR solver from openQCD~\cite{openQCD}. Contractions were
performed with a high-performance BLAS library using the Python
package opt\_einsum~\cite{opt_einsum}.  The correlator analysis was
done using SigMonD~\cite{sigmond}.  Much of the data handling and the
subsequent phase shift analysis was done using NumPy~\cite{numpy} and
SciPy~\cite{scipy}. The plots were prepared using
Matplotlib~\cite{Hunter:2007}. The quantization condition was computed
using TwoHadronsInBox~\cite{Morningstar:2017spu}.
This research was partly supported by Deutsche Forschungsgemeinschaft
(DFG, German Research Foundation) through the Cluster of Excellence
``Precision Physics, Fundamental Interactions and Structure of
Matter'' (PRISMA+ EXC 2118/1) funded by the DFG within the German
Excellence Strategy (Project ID 39083149), as well as the
Collaborative Research Centers SFB 1044 ``The low-energy frontier of
the Standard Model'' and CRC-TR 211 ``Strong-interaction matter under
extreme conditions'' (Project ID 315477589 -- TRR 211).  ADH is
supported by: (i) The U.S. Department of Energy, Office of Science,
Office of Nuclear Physics through the Contract No. DE-SC0012704
(S.M.); (ii) The U.S. Department of Energy, Office of Science, Office
of Nuclear Physics and Office of Advanced Scientific Computing
Research, within the framework of Scientific Discovery through Advanced
Computing (SciDAC) award Computing the Properties of Matter with
Leadership Computing Resources. JRG acknowledges support from the
Simons Foundation through the Simons Bridge for Postdoctoral
Fellowships scheme. We are grateful to our colleagues within the CLS
initiative for sharing ensembles.

\bibliographystyle{JHEP}
\bibliography{refs}

\end{document}